\documentclass[%
 aip,
 jmp,%
 amsmath,amssymb,
 preprint,%
]{revtex4-1}

\usepackage{graphicx}
\usepackage{dcolumn}
\usepackage{bm}

\begin{document}

\preprint{AIP/123-QED}

\title{Current-induced phase transition in ballistic Ni nanocontacts}%

\author{R.G. Gatiyatov }%
 \email{Ruslan.Gatiyatov@gmail.com}
 \affiliation{Zavoisky Physical-Technical Institute RAS, 420029 Kazan, Russia}%
\author{V.N. Lisin}
\author{A.A. Bukharaev}
\date{\today}
\begin{abstract}
Local phase transition from ferromagnetic to paramagnetic state in the region of the ballistic Ni nanocontacts (NCs) has been  experimentally observed. We found that contact size reduction leads to an increase in the bias voltage at which the local phase transition occurs. Presented theoretical interpretation of this phenomena takes into the account the specificity of the local heating of the ballistic NC and describes the electron's energy relaxation dependences on the applied voltage. The experimental data are in good qualitative and quantitative agreement with the theory proposed.
\end{abstract}

\maketitle
For more than a decade a point contact between two metal conductors with characteristic micro- and nanoscale dimensions is one of the interesting objects studied theoretically and experimentally \cite{Yanson,Agrait}. Recently, the investigations were mainly concentrated on the atomic scale contacts due to the rich variety of the quantum size effects observed in such systems \cite{Chopra, Huntington, Calvo}. However, the nanocontacts (NCs) with the transverse dimensions of the order of an electron Fermi wavelength $\lambda_{F}$ are not suitable for practical usage because of their rapid destruction. Therefore, ballistic NCs with the diameter $d$ larger than $\lambda_{F}$, on the one hand, but smaller or comparable to the transport electron mean free path $l_{tr}$, on the other hand, are more promising for applications. The investigation of the specificity of heating of the magnetic ballistic NCs and the adjacent regions by electrons current is one of the topics that is particularly due to the spin transfer torque effect in such structures \cite{Ralph} where temperature can play a crucial role \cite{Hatami}. Moreover, the heating of the contact region to the Curie temperature should lead to a phase transition from ferromagnetic to paramagnetic state. Earlier, such phase transition was achieved in a number of ferromagnetic microcontacts ($l_{tr}<<d$) by current-induced heating at liquid helium temperatures \cite{Verkin}. However, up to now the phase transition of the ferromagnetic ballistic NCs was not investigated. In this case ($d\simeq l_{tr}$), it is known that the voltage drops in the region of the order of the contact diameter and the electrons release their excess energy beyond the the region of the potential drop \cite{Levinson}. It determines the specificity of the thermal heating of the contact and the adjacent regions that is of interest of this research.

Ni NCs has been formed  between two microwires fixed on a substrate using electrochemical method \cite{Gatiyatov}. NCs were fabricated in the nickel sulphate solution $0.25$\,M NiSO$_4$ + $0.5$\,M H$_3$BO$_3$ (working voltage $1$-$1.4$\,V). The conductance and the current-voltage (I-V) characteristics were recorded with two digital multimeters Agilent 34410A using 4-probe method. A single 70 Hz triangular voltage pulse was applied to the circuit to obtain current-voltage curves.  I-V curves were recorded at room temperature in the bath with bidistilled water, which has the conductivity much smaller than a NC conductivity. The zero-bias resistance of the fabricated Ni NCs was in the range of $30$-$400$\,Ohm.

\begin{figure}
	\includegraphics{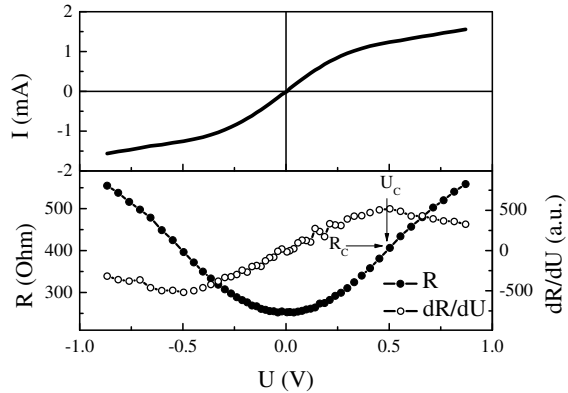}
		\caption{\label{fig:fig1} I-V curve (top) and typical dependence of the resistance and $dR/dU$ spectrum of the Ni NC versus applied voltage (bottom).}
\end{figure}

A typical I-V curve and the dependences of the $dR/dU$ spectrum and the resistance of the Ni NC versus applied voltage are shown in Fig.~\ref{fig:fig1}. Forward and backward branches of the current-voltage curve coincide. Therefore, we conclude that the current does not affect the contact, and softening effect \cite{Holm} or influence of the solution \cite{Gatiyatov} are absent or negligible in our case.

\begin{figure}
	\includegraphics{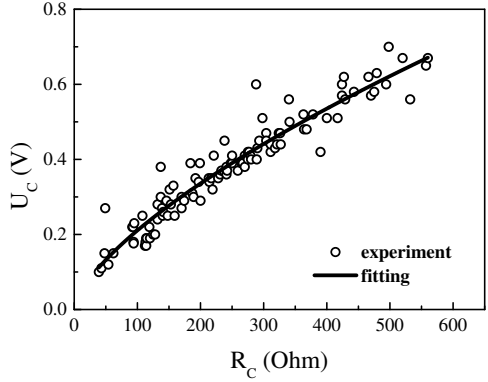}
		\caption{\label{fig:fig2} 
	Dependence of the bias voltage $U_C$ at which the local phase transition occurs on the resistance $R_C$ of the Ni NCs at this voltage (see Fig.~\ref{fig:fig1}). Each dot was obtained from the individual $dR/dU$ spectrum of the Ni NC with different zero-bias resistance. Solid line is the result of the fitting by Eq.(\ref{eq:eq7}).}
\end{figure}

I-V curves are nonlinear and the resistance increases with the increase in the applied voltage.
Moreover, the $dR/dU$ spectra have a maximum. The position of the maximum depends on the resistance (size) of the Ni NC (Fig.~\ref{fig:fig2}). It is worth noting that such experiments were also carried out for Cu NCs. In case of Cu the $dR/dU$ spectra do not contain any features in the same region of the applied voltages.

The observed feature on the $dR/dU$ spectra we attribute to the local phase transition from ferromagnetic to paramagnetic state in the vicinity of the Ni NC region. Above the Curie temperature ($T_C=631$\,K for Ni) the electron-magnon scattering saturates and becomes temperature-independent \cite{Vonsovsky}. So, $dR/dU$ is determined only by electron-phonon scattering. It should lead to the maximum in the $dR/dU$ spectrum versus $U$ that was observed experimentally (Fig.~\ref{fig:fig1}).

From the solution of the steady-state heat conduction equation one can find the contact resistance $R_C$ at critical voltage $U_C$ at which the local phase transition occurs . Let us make several assumptions. First of all, let us assume that the form of the NC matches with an orifice that is the diameter of the NC $d$ is much larger than its length. Secondly, let's assume that the relaxation of the electron's excess energy occurs in a sphere with the radius $b$ around the contact. Assuming  the density of the released thermal power constant, the problem becomes spherically symmetrical. 

In the steady state, the thermal flux through the sphere with the radius $r$ is equal to the released power in this sphere:
\begin{eqnarray}\label{eq:eq1}
    \oint\limits_{S_r} { \textbf{q}\,d\textbf{S}} = \left\{
        \begin{array}{ll}
            \frac{U^2}{R}\frac{r^3}{b^3} &\mbox{ $r<b$} \\
            \frac{U^2}{R} &\mbox{ $r\geq b$}
        \end{array} \right.,
\end{eqnarray}
where $\textbf{q}=-\lambda\nabla T$ is the thermal flux, $\lambda$ is the volume thermal conductivity, $U$ is the voltage applied to the NC, $R$ is the NC resistance, $S_r$ is the surface of the sphere with radius $r$.
By solving the system of the equations (\ref{eq:eq1}) one can obtain the relation between the temperature $T_b$ on the sphere with the radius $b$ and the temperature $T_{nc}$ in the center of the NC ($r=0$):
\begin{eqnarray}\label{eq:eq2}
    \int\limits_{T_0}^{T_b}\lambda \left( T'\right) dT'=\frac{1}{4\pi b} \frac{U^2}{R}= \frac{2}{3}
    \int\limits_{T_0}^{T_{nc}}\lambda \left( T'\right) dT',
\end{eqnarray}
where $T_0$ is the the temperature of the conductors far away from the NC.

The $b$ value is determined \cite{Levinson} by the dimension of the region where the external potential drops (of the order of the NC diameter \cite{Agrait, Levinson}) and the electron diffusion length $\Lambda_E$ over the energy relaxation time $\tau_{E}$ and can be estimated as:
\begin{eqnarray}\label{eq:eq3}
	 b = d + \Lambda_{E},
\end{eqnarray}
where
\begin{eqnarray}\label{eq:eq4}
	\Lambda_{E}=\sqrt{6 D_{tr}\left(T\right)\tau_E\left(T, U\right)} = \sqrt{2 l_{tr}(T) l_E(T, U)}.
\end{eqnarray}
Here $D_{tr} =\frac{1}{3} \upsilon_F l_{tr} $ is the transport diffusion coefficient of the electrons, $\upsilon_F$ is the Fermi velocity, $l_{tr}=\tau_{tr}\upsilon_F$, $l_E=\tau_E\upsilon_F$, $\tau_{tr}$ is the transport relaxation time.

Now we take a look at the mechanisms of the momentum and energy-loss mechanisms of the hot electrons in more detail. In ferromagnetic metals, the excess energy $E$ of the electrons can be transferred either to the lattice via the phonon emission or to the magnetic subsystem via the magnon emission. If the excess energy of the electron $E$ strongly exceeds the energy of the phonons with the Debye frequency ($E_D=35$\,meV for Ni \cite{Verkin}), then at each act of the phonon emission an electron radiates a phonon with the Debye frequency and undergoes large-angle scattering \cite{Gantmaher}. Therefore, only one act of the scattering is needed for the electron in order to forget the direction of its motion. The mean time between such emissions $\tau_{eph}$ determines the electron-phonon contribution to the transport relaxation time $\tau_{tr}$. For the time $t$ the electron emits $t/t_{eph}$ phonons with the Debye frequency and looses energy $E_D t/\tau_{eph}$. In the case of the electron-magnon scattering, the mean time between emissions of the magnons $\tau_{em}$ plays the role of $\tau_{eph}$ and the exchange energy $E_{ex}$ ($E_{ex}=54.5$\,meV for Ni \cite{Vonsovsky}) plays the role of $E_D$. Then the effective time $\tau_{E}$ needed the electron to entirely lose the excess energy $E$ is proportional to this energy:
\begin{eqnarray}\label{eq:eq5}
	\tau_{E} = \frac{E}{E_D/\tau_{eph} + E_{ex} /\tau_{em}}.
\end{eqnarray}
From Eq.~(\ref{eq:eq5}) one can see that for $E >> E_D,E_{ex}$ the energy relaxation time $\tau_E$ becomes much larger than the transport time $\tau_{tr}$ (neglecting electron-impurity scattering $\tau_{tr}^{-1} = \tau_{eph}^{-1} + \tau_{em}^{-1}$ ).

Let us now consider the case $d < l_{tr}$. In this regime, the electron ballistically passes the region where most of the applied voltage drops and it gains the energy $E = |eU|$ ($e$ is the electron charge). By substituting Eq.~(\ref{eq:eq5}) into Eq.~(\ref{eq:eq3}) and taking into the account that $d << \Lambda_E$, the expression for $b$ is:
\begin{eqnarray}\label{eq:eq6}
    b(\tilde{T},U) \simeq \sqrt{
    \frac{2 l_{tr} (\tilde{T}) |eU|}{E_D/l_{eph}(\tilde{T}) +
    E_{ex}/l_{em}(\tilde{T})}}\propto \sqrt{|U|},
\end{eqnarray}
where $\tilde{T}$ is the temperature averaged over a sphere with radius $b$ (the temperature is not constant inside the sphere), $l_{eph}=\tau_{eph}\upsilon_F$, $l_{em}=\tau_{em}\upsilon_F$. Therefore, according to Eq.~(\ref{eq:eq6}), the size of the heating region $b$ depends on the applied external potential difference. This effect separates the ballistic NCs from the microcontacts and determines the specificity of the current-induced heating.

By substituting Eq.~(\ref{eq:eq6}) into Eq.~(\ref{eq:eq2}), one can obtain the dependence of the voltage $U_C$ needed to heat the NC to the critical temperature $T_C$ on the NC resistance $R_C$:
\begin{eqnarray}\label{eq:eq7}
    U_C = A R_C^{\frac{2}{3}},
\end{eqnarray}
where $A$ is constant:
\begin{eqnarray}\label{eq:eq8}
    A = \left[ \frac{8 \pi}{3} \sqrt{
    \frac{2 l_{tr} (\tilde{T}) |e|}{\frac{E_D}{l_{eph}(\tilde{T})} +
    \frac{E_{ex}}{l_{em}(\tilde{T})}}}
\int\limits_{T_0}^{T_C} \lambda (T')\,dT'\right]^{ \frac{2}{3} }.
\end{eqnarray}

Let us now demonstrate that Eq.~(\ref{eq:eq7}) can be applied to the experimental dependence $U_C(R_C)$ i.e., the condition $d < l_{tr}(T_C)$ holds. The diameter of the fabricated Ni NC at $T_0=300$\,K can be found using the value of its zero-bias resistance from Wexler formula \cite{Wexler}. To this end, one needs to know $l_{tr}(T_0)$ and the value of the electrical resistivity $\rho(T_0)=7.2 \cdot 10^{-8}$\,Ohm$\cdot$m \cite{Handbook} for Ni. We estimated the values of the mean free paths as $l(T)=5 \cdot 10^{-16}/\rho(T)$.  This gives $l_{tr}(T_0)=7$\,nm and $l_{tr}(T_C)=2$\,nm. According to Wexler formula, the NCs with the zero-bias resistance of $30$, $100$ and $400$\,Ohm have the diameters of $6.4$, $3.2$ and $1.5$\,nm, respectively. Therefore, the ballistic transport of electrons ($d<l_{tr}(T_C)$) should exist for the contacts with $R>100$\,Ohm even at the Curie temperature.

Another important condition of the theory proposed is the constancy of the phase transition temperature for the contacts of different size.We suppose that the surface and size effects are negligible and can not strongly affect the value of the Curie temperature of the contact region. This means the phase transition occurs at the same temperature for the nanocontacts of different size.

A solid curve in Fig.~\ref{fig:fig2} represents the result of the fitting of the experimental data using the function $U_C=A R_C^{\alpha}$ by root-mean-square method.

The exponential value $\alpha=0.67\pm0.02$, which is in good agreement with the value from Eq.~(\ref{eq:eq7}), and the coefficient $A=(0.95\pm0.12)\cdot10^{-2}$\,V/Ohm$^\frac{2}{3}$ were extracted using fitting. One can see that the exponential law Eq.~(\ref{eq:eq7}) describes well the experimental dependence $U_C(R_C)$ for $|e U_C| > E_D,E_{ex}$ in wide range of the NC resistances. The estimate of the $A$ using Eq.~(\ref{eq:eq8}) gives the value of $2.2\cdot10^{-2}$\,V/Ohm$^\frac{2}{3}$, which is of the same order of magnitude as the experimental one. Here we used the parameter values $\tilde{T} = (T_C+T_b)/2 = 570$\,K ($T_b = 508$\,K was found from Eq.~(\ref{eq:eq2}) for $T_{nc}=T_C$ and the known dependence $\lambda(T)$ \cite{Handbook}), $l_{tr}(\tilde{T})=2.6$\,nm, $l_{eph}(\tilde{T})=3.9$\,nm and $l_{em}(\tilde{T})=8.2$\,nm. The values of the mean free paths were estimated as $l(T)=5 \cdot 10^{-16}/\rho(T)$ where the $\rho_{eph}$, $\rho_{em}$ and $\rho_{tr}=\rho_{eph}+\rho_{em}$ values for Ni were taken from \cite{Handbook}.

For the Ni NC with the smallest zero-bias resistance of $30$\,Ohm ($R_C=40$\,Ohm) the electron transport is diffusive because $d>l_{tr}(T_C)$. The maximum on the $dR/dU$ spectrum is observed at $0.12 \pm 0.02$\,V. That is in good agreement with the voltage of $0.19$\,V needed to heat the diffusive Ni microcontact to the critical temperature \cite{Verkin}. Here one has to take into the account that in \cite{Verkin} the heating was carried out at the liquid helium temperature.

The estimates are in good qualitative and quantitative agreement with the experimental results. It confirms that the observed feature on the $dR/dU$ spectra is due to the phase transition from ferromagnetic to paramagnetic state in the region of the ballistic contact. 

The main specificity of the current-induced heating of the ballistic NCs is in the voltage dependence of the heated region size for the high applied voltages ($|e U|>>E_D, E_{ex}$). It results in the increase in the bias voltage needed to heat NC to the critical temperature (Fig.~\ref{fig:fig2}) with the reduce of the Ni contact's size (the increase in the NC resistance). While in microcontacts it does not depend on the contact size \cite{Verkin}.

It must be noted that  I-V curves of the ballistic Ni NCs were studied in the past \cite{Sullivan} for $U<<U_C$, where only a negligible deviation from the linearity was found. The authors concluded \cite{Chopra, Sullivan}  that I-V curves of clean metallic ballistic Ni NCs must be linear and the origin of the nonlinearity is due to the contamination of the contact \cite{Hansen}. In our case (fig.2) the nonlinearity of the I-V curves is also negligible at small applied voltages as in \cite{Sullivan}. While the nonlinearity becomes noticeable at high voltage ($U \sim U_C$). $U_C$ increases with the reduce of the NC size. In case of contaminated NCs the resistance drops with an increase of the applied voltage \cite{Hansen} that is in contrast to our results. We think the influence of the contaminations can be eliminated in our case.

In conclusion, we have shown that I-V curves of the ballistic metallic Ni NCs ($d \leq l_{tr}$) are nonlinear in the region of high applied voltages ($U \sim U_C$) as a result of Joule heating of the contact region. Such heating results in the phase transition in the contact region if the critical temperature reached.

We are grateful to S.A. Ziganshina for her help in preparation of chemical solutions. This work was supported by the grant of the Russian Foundation for Basic Research No. $09$-$02$-$00568$ and by the programs of the Division of Physical Sciences of RAS.

\nocite{*}
\providecommand{\noopsort}[1]{}\providecommand{\singleletter}[1]{#1}%

\begin{thebibliography}{10}%
\makeatletter
\providecommand \@ifxundefined [1]{%
 \ifx #1\undefined \expandafter \@firstoftwo
 \else \expandafter \@secondoftwo
\fi
}%
\providecommand \@ifnum [1]{%
 \ifnum #1\expandafter \@firstoftwo
 \else \expandafter \@secondoftwo
\fi
}%
\providecommand \enquote [1]{``#1''}%
\providecommand \bibnamefont  [1]{#1}%
\providecommand \bibfnamefont [1]{#1}%
\providecommand \citenamefont [1]{#1}%
\providecommand\href[0]{\@sanitize\@href}%
\providecommand\@href[1]{\endgroup\@@startlink{#1}\endgroup\@@href}%
\providecommand\@@href[1]{#1\@@endlink}%
\providecommand \@sanitize [0]{\begingroup\catcode`\&12\catcode`\#12\relax}%
\@ifxundefined \pdfoutput {\@firstoftwo}{%
 \@ifnum{\z@=\pdfoutput}{\@firstoftwo}{\@secondoftwo}%
}{%
 \providecommand\@@startlink[1]{\leavevmode}%
 \providecommand\@@endlink[0]{}%
}{%
 \providecommand\@@startlink[1]{%
  \leavevmode
  \pdfstartlink
   attr{/Border[0 0 1 ]/H/I/C[0 1 1]}%
   user{/Subtype/Link/A<</Type/Action/S/URI/URI(#1)>>}%
  \relax
 }%
 \providecommand\@@endlink[0]{\pdfendlink}%
}%
\providecommand \url  [0]{\begingroup\@sanitize \@url }%
\providecommand \@url [1]{\endgroup\@href {#1}{\urlprefix}}%
\providecommand \urlprefix [0]{URL }%
\providecommand \Eprint[0]{\href }%
\@ifxundefined \urlstyle {%
  \providecommand \doi [1]{doi:\discretionary{}{}{}#1}%
}{%
  \providecommand \doi [0]{doi:\discretionary{}{}{}\begingroup
  \urlstyle{rm}\Url }%
}%
\providecommand \doibase [0]{http://dx.doi.org/}%
\providecommand \Doi[1]{\href{\doibase#1}}%
\providecommand \selectlanguage [0]{\@gobble}%
\providecommand \bibinfo [0]{\@secondoftwo}%
\providecommand \bibfield [0]{\@secondoftwo}%
\providecommand \translation [1]{[#1]}%
\providecommand \BibitemOpen[0]{}%
\providecommand \bibitemStop [0]{}%
\providecommand \bibitemNoStop [0]{.\EOS\space}%
\providecommand \EOS [0]{\spacefactor3000\relax}%
\providecommand \BibitemShut [1]{\csname bibitem#1\endcsname}%
\bibitem{Yanson}%
  \BibitemOpen
  \bibfield{author}{%
  \bibinfo {author} {\bibfnamefont{A.~G.~M.}\ \bibnamefont{Jansen}}, \bibinfo
  {author} {\bibfnamefont{A.~P.}\ \bibnamefont{van Gelder}},\ and\ \bibinfo
  {author} {\bibfnamefont{P.}~\bibnamefont{Wyder}},\ }%
  \bibfield{journal}{%
  \bibinfo {journal} {J.\ Phys.\ C:\ Solid State Phys.}\ }%
  \textbf{\bibinfo {volume} {13}},\ \bibinfo {pages} {6073} (\bibinfo {year}
  {1980})\BibitemShut{NoStop}%
\bibitem{Agrait}%
  \BibitemOpen
  \bibfield{author}{%
  \bibinfo {author} {\bibfnamefont{N.}~\bibnamefont{Agrait}}, \bibinfo {author}
  {\bibfnamefont{A.~L.}\ \bibnamefont{Yeyati}},\ and\ \bibinfo {author}
  {\bibfnamefont{J.~M.}\ \bibnamefont{van Ruitenbeek}},\ }%
  \bibfield{journal}{%
  \bibinfo {journal} {Phys.\ Rep.}\ }%
  \textbf{\bibinfo {volume} {377}},\ \bibinfo {pages} {81} (\bibinfo {year}
  {2003})\BibitemShut{NoStop}%
\bibitem{Chopra}%
  \BibitemOpen
  \bibfield{author}{%
  \bibinfo {author} {\bibfnamefont{H.~D.}\ \bibnamefont{Chopra}}, \bibinfo
  {author} {\bibfnamefont{M.~R.}\ \bibnamefont{Sullivan}}, \bibinfo {author}
  {\bibfnamefont{J.~N.}\ \bibnamefont{Armstrong}},\ and\ \bibinfo {author}
  {\bibfnamefont{S.~Z.}\ \bibnamefont{Hua}},\ }%
  \bibfield{journal}{%
  \bibinfo {journal} {Nature Materials}\ }%
  \textbf{\bibinfo {volume} {4}},\ \bibinfo {pages} {832} (\bibinfo {year}
  {2005})\BibitemShut{NoStop}%
\bibitem{Huntington}%
  \BibitemOpen
  \bibfield{author}{%
  \bibinfo {author} {\bibfnamefont{M.~D.}\ \bibnamefont{Huntington}}, \bibinfo
  {author} {\bibfnamefont{M.~R.~Sullivan}\ \bibnamefont{J.~N.~Armstrong}},
  \bibinfo {author} {\bibfnamefont{S.~Z.}\ \bibnamefont{Hua}},\ and\ \bibinfo
  {author} {\bibfnamefont{H.~D.}\ \bibnamefont{Chopra}},\ }%
  \bibfield{journal}{%
  \bibinfo {journal} {Phys.\ Rev.\ B}\ }%
  \textbf{\bibinfo {volume} {78}},\ \bibinfo {pages} {035442} (\bibinfo {year}
  {2008})\BibitemShut{NoStop}%
\bibitem{Calvo}%
  \BibitemOpen
  \bibfield{author}{%
  \bibinfo {author} {\bibfnamefont{M.~R.}\ \bibnamefont{Calvo}}, \bibinfo
  {author} {\bibfnamefont{J.}~\bibnamefont{Fernandez-Rossier}}, \bibinfo
  {author} {\bibfnamefont{J.~J.}~\bibnamefont{Palacios}}, \bibinfo
  {author} {\bibfnamefont{D.}~\bibnamefont{Jacob}}, \bibinfo
  {author} {\bibfnamefont{D.}~\bibnamefont{Natelson}}, \ and\ \bibinfo
  {author} {\bibfnamefont{C.}\ \bibnamefont{Untiedt}},\ }%
  \bibfield{journal}{%
  \bibinfo {journal} {Nature}\ }%
  \textbf{\bibinfo {volume} {458}},\ \bibinfo {pages} {1150} (\bibinfo {year}
  {2009})\BibitemShut{NoStop}%
\bibitem{Ralph}%
  \BibitemOpen
  \bibfield{author}{%
  \bibinfo {author} {\bibfnamefont{D.~C.}\ \bibnamefont{Ralph}}\ and\ \bibinfo
  {author} {\bibfnamefont{M.~D.}\ \bibnamefont{Stiles}},\ }%
  \bibfield{journal}{%
  \bibinfo {journal} {J.\ Magn.\ Mag.\ Mat.}\ }%
  \textbf{\bibinfo {volume} {320}},\ \bibinfo {pages} {1190} (\bibinfo {year}
  {2008})\BibitemShut{NoStop}%
\bibitem{Hatami}%
  \BibitemOpen
  \bibfield{author}{%
  \bibinfo {author} {\bibfnamefont{M.}~\bibnamefont{Hatami}}, \bibinfo {author}
  {\bibfnamefont{G.~E.~W.}\ \bibnamefont{Bauer}}, \bibinfo {author}
  {\bibfnamefont{Q.}~\bibnamefont{Zhang}},\ and\ \bibinfo {author}
  {\bibfnamefont{P.~J.}\ \bibnamefont{Kelly}},\ }%
  \bibfield{journal}{%
  \bibinfo {journal} {Phys.\ Rev.\ Lett.}\ }%
  \textbf{\bibinfo {volume} {99}},\ \bibinfo {pages} {066603} (\bibinfo {year}
  {2007})\BibitemShut{NoStop}%
\bibitem{Verkin}%
  \BibitemOpen
  \bibfield{author}{%
  \bibinfo {author} {\bibfnamefont{B.~I.}\ \bibnamefont{Verkin}}, \bibinfo
  {author} {\bibfnamefont{J.~K.}\ \bibnamefont{Yanson}}, \bibinfo {author}
  {\bibfnamefont{I.~K.}\ \bibnamefont{Kulik}}, \bibinfo {author}
  {\bibfnamefont{O.~I.}\ \bibnamefont{Shklyarevski}}, \bibinfo {author}
  {\bibfnamefont{A.~A.}\ \bibnamefont{Lysykh}},\ and\ \bibinfo {author}
  {\bibfnamefont{Yu.~G.}\ \bibnamefont{Nayduk}},\ }%
  \bibfield{journal}{%
  \bibinfo {journal} {Solid State Commun.}\ }%
  \textbf{\bibinfo {volume} {30}},\ \bibinfo {pages} {215} (\bibinfo {year}
  {1979})\BibitemShut{NoStop}%
\bibitem{Levinson}%
  \BibitemOpen
  \bibfield{author}{%
  \bibinfo {author} {\bibfnamefont{M.}~\bibnamefont{Rokni}}\ and\ \bibinfo
  {author} {\bibfnamefont{Y.}~\bibnamefont{Levinson}},\ }%
  \bibfield{journal}{%
  \bibinfo {journal} {Phys.\ Rev.\ B}\ }%
  \textbf{\bibinfo {volume} {52}},\ \bibinfo {pages} {1882} (\bibinfo {year}
  {1995})\BibitemShut{NoStop}%
\bibitem{Gatiyatov}%
  \BibitemOpen
  \bibfield{author}{%
  \bibinfo {author} {\bibfnamefont{R.~G.}\ \bibnamefont{Gatiyatov}}, \bibinfo
  {author} {\bibfnamefont{S.~A.}\ \bibnamefont{Ziganshina}}, ,\ and\ \bibinfo
  {author} {\bibfnamefont{A.~A.}\ \bibnamefont{Bukharaev}},\ }%
  \bibfield{journal}{%
  \bibinfo {journal} {JETP Lett.}\ }%
  \textbf{\bibinfo {volume} {86}},\ \bibinfo {pages} {412} (\bibinfo {year}
  {2007})\BibitemShut{NoStop}%
 \bibitem{Holm}%
  \BibitemOpen
  \bibfield{author}{%
  \bibinfo {author} {\bibfnamefont{R.}\ \bibnamefont{Holm}},\ }%
  \emph{\bibinfo {title} {Electrical contacts [Russian Translation]}},\ (\bibinfo {publisher} {Izd. Inostr. Lit., Moscow},\
  \bibinfo {year} {1961})\BibitemShut{NoStop}%
\bibitem{Vonsovsky}%
  \BibitemOpen
  \bibfield{author}{%
  \bibinfo {author} {\bibfnamefont{S.V.}\ \bibnamefont{Vonsovsky}},\ }%
  \emph{\bibinfo {title} {Magnetism}},\ (\bibinfo {publisher} {Nauka, Moscow},\
  \bibinfo {year} {1971})\BibitemShut{NoStop}%
\bibitem{Gantmaher}%
  \BibitemOpen
  \bibfield{author}{%
  \bibinfo {author} {\bibfnamefont{V.~F.}\ \bibnamefont{Gantmaher}}\ and\
  \bibinfo {author} {\bibfnamefont{I.~B.}\ \bibnamefont{Levinson}},\ }%
  \emph{\bibinfo {title} {Carrier Scattering in Metals and Semiconductors}},\
  \bibinfo {series} {Modern Problems in Condensed Matter Science},
  Vol.~\bibinfo {volume} {19}\ (\bibinfo {publisher} {Nauka, Moscow},\ \bibinfo
  {year} {1984})\BibitemShut{NoStop}%
\bibitem{Wexler}%
  \BibitemOpen
  \bibfield{author}{%
  \bibinfo {author} {\bibfnamefont{G.}~\bibnamefont{Wexler}},\ }%
  \bibfield{journal}{%
  \bibinfo {journal} {Proc.\ Phys.\ Soc.}\ }%
  \textbf{\bibinfo {volume} {89}},\ \bibinfo {pages} {927} (\bibinfo {year}
  {1966})\BibitemShut{NoStop}%
\bibitem{Handbook}%
  \BibitemOpen
  \bibfield{author}{%
  \bibinfo {author} {\bibfnamefont{D.~R.}\ \bibnamefont{Lide}},\ }%
  \emph{\bibinfo {title} {Handbook of chemistry and physics 84-th edition}}\
  (\bibinfo {publisher} {CRC press, Boca Raton, Florida},\ \bibinfo {year}
  {2003})\BibitemShut{NoStop}%
\bibitem{Sullivan}%
  \BibitemOpen
  \bibfield{author}{%
  \bibinfo {author} {\bibfnamefont{M.~R.}\ \bibnamefont{Sullivan}}, \bibinfo
  {author} {\bibfnamefont{D.~A.}\ \bibnamefont{Boehm}}, \bibinfo {author}
  {\bibfnamefont{D.~A.}\ \bibnamefont{Ateya}},\bibinfo {author}
  {\bibfnamefont{S.~Z.}\ \bibnamefont{Hua}},\ and\ \bibinfo {author}
  {\bibfnamefont{H.~D.}\ \bibnamefont{Chopra}},\ }%
  \bibfield{journal}{%
  \bibinfo {journal} {Phys.\ Rev.\ B}\ }%
  \textbf{\bibinfo {volume} {71}},\ \bibinfo {pages} {024412} (\bibinfo {year}
  {2005})\BibitemShut{NoStop}%
\bibitem{Hansen}%
  \BibitemOpen
  \bibfield{author}{%
  \bibinfo {author} {\bibfnamefont{K.}\ \bibnamefont{Hansen}}, \bibinfo
  {author} {\bibfnamefont{S.~K.}\ \bibnamefont{Nielsen}}, \bibinfo {author}
  {\bibfnamefont{M.}\ \bibnamefont{Brandbyge}},\bibinfo {author}
  {\bibfnamefont{E.}\ \bibnamefont{Lagsgaard}},\bibinfo {author}
  {\bibfnamefont{I.}\ \bibnamefont{Stensgaard}},\ and\ \bibinfo {author}
  {\bibfnamefont{F.}\ \bibnamefont{Besenbacher}},\ }%
  \bibfield{journal}{%
  \bibinfo {journal} {Appl.\ Phys.\ Lett.}\ }%
  \textbf{\bibinfo {volume} {77}},\ \bibinfo {pages} {708} (\bibinfo {year}
  {2000})\BibitemShut{NoStop}%
\end{thebibliography}
\end{document}